# Static magnetic field stimulates growth of maize seeds

L. M. Ferroni[1, #], M. I. Dolz[2,3, #], M. F. Guerra[3], L. Makinistian[2,3,*]

[1] Instituto Nacional de Tecnología Agropecuaria, Estación Experimental Villa Mercedes, San Luis, Argentina

[2] Department of Physics, Universidad Nacional de San Luis (UNSL), San Luis, Argentina

[3] Instituto de Física Aplicada (INFAP), Universidad Nacional de San Luis (UNSL)-CONICET, San Luis, Argentina

# These authors contributed equally to this work.

* Correspondence to: L. Makinistian, Department of Physics and Instituto de Física Aplicada (INFAP), Universidad Nacional de San Luis-CONICET, San Luis, Argentina, Ejército de los Andes 950, CP5700, San luis, Argentina. Email: lmakinistian@unsl.edu.ar Telephone: +54 266 452 0329

## Abstract

The physical pre-treatment of seeds is a growing field of research in agricultural biotechnology. Among its possibilities, magnetopriming of seeds has received increasing attention in the last two decades. Inspired by remarkable reports of the literature on the effects of static magnetic fields (SMF) on maize seeds, we performed similar experiments, though expanding the range of SMF up to 350 mT, higher than most of the reported studies. With exposure durations of 1 h, we tested 7 different SMF intensities, from 50 to 350 mT at increments of 50 mT. We challenged our findings with an exhaustive analysis of the background static and alternated magnetic fields inside the stove where the germinations took place, in order to rule out background fields as a confounding variable. We found a maximum effect of 108.9 % increase (more than double) in the average total length of plantules at 150 mT, at the 10th day of germination. All other intensities, except 350 mT, also induced a significant growth stimulation. While not seeming to represent a determinant factor, our analysis calls the attention to the possible relevance of different conditions in different levels and zones of standard stoves. Our findings are in line with numerous studies pointing to magnetopriming of maize seeds as a novel, viable and reproducible physical treatment for the enhancement of germination.

## Statements and Declarations

**Competing Interests:** The authors declare no competing interests.

**Acknowledgements:** The authors thank Dr. Federico Romá for valuable discussions on the design and execution of this study. Grant sponsors: Universidad Nacional de San Luis, Argentina (PROICO 02-0518 and PROICO 03-2220).

**Authors contributions:** LMF: Conceptualization and germination assays; MID: Conceptualization and exposure of seeds to magnetic fields; MFG: Data curation, statistical analysis and figures; LM: Conceptualization, data curation, statistical analysis and figures, and writing-first draft. All authors reviewed the manuscript and approved it for submission.

# Introduction

The physical treatment of seeds for their invigoration is an active and rapidly progressing area of agricultural biotechnology. This approach aims at the enhancement of crop establishment, germination, and yield by exposing seeds to diverse physical stimuli (Araújo et al. 2016; Farooq et al. 2019). While still not massively adopted, a priori, these methods promise to be cost-effective and eco-friendly in that they do not involve any chemical substances; hence, reducing costs of production and avoiding well-known risks of toxicity for workers and populations near the crop fields. Among these physical methods, exposure to static magnetic fields (SMF) could turn out advantageous as compared to microwave and ionizing radiations (UV-, X- and γ-rays) due to the relative simplicity, lower cost and higher safety of the technology needed for its implementation (i.e., permanent magnets and/or electromagnets). The treatment of seeds with SMF, so called "magnetopriming", has been abundantly reported to have beneficial effects in plant physiology, enhancing germination rates, root development, vigour and seedling biomass, photosynthetic pigment content, water and nutrients uptake, and response to biotic and abiotic stressing factors, e.g., tomato yellow leaf curl virus (TYLCV), salt, drought, or heavy metal contamination in soil (De Souza et al. 2006; Anand et al. 2012; Radhakrishnan 2019; Sarraf et al. 2020).

While the question about the bio-physico-chemical mechanisms that underlie these effects remains broadly open (we further elaborate on this point below), there is also the more pragmatic quest of searching, for each crop, the specific parameters of exposure that would maximize the beneficial effects on the seeds. Rather than whether magnetopriming is possible, the pressing question is how to optimize it. Given the current lack of rationale that would predict optimal exposure, the state of the art rules that screening of multiple parameters is for the time being the best approach. The domain of exploration includes typical field intensities in the order of tens and hundreds of mT (milliTeslas) and exposure durations raging from seconds (Atak et al. 2003), to minutes (Zepeda-Bautista et al. 2014) and hours (Florez et al. 2007).

Some of the crops that have been reported to benefit from exposure to SMF are sweet pepper (Ahamed et al. 2013), soybean (Shine et al. 2011), wheat (Rathod and Anand 2016), barley (Hozayn and Ahmed 2019), and chickpea (Thomas et al. 2013), among many others, including maize (Zea mays L.). This cereal is one of the most important crops worldwide.

According to the 2020 Statistical yearbook of the World Food and Agriculture Organization of the United Nations (FAO 2020) (p. 11), in 2018 four crops accounted for half of the total worldwide production: sugar cane (21 %, 1.9 billon tonnes), maize (13 %, 1.1 billon tonnes), rice (9 %, 0.8 billon tonnes) and wheat (8 %, 0.7 billon tonnes). Among various studies of magnetopriming of maize seeds (Aladjadjiyan 2002; Domínguez-Pacheco et al. 2010; Anand et al. 2012; Shine et al. 2012; Martinez et al. 2017; Torres et al. 2018, 2019), one from 2017 (Vashisth and Joshi 2017) outstands for the exhaustiveness of the screening the authors performed. They assessed several endpoints (percentage germination, speed of germination, seedling length, and seedling dry weight) for several exposures: 50, 100, 150, 200, and 250 mT, each of them for 1, 2, 3, and 4 h. Their conclusion was clear: in general, the MF improved the growth of the seeds and, in particular, the most effective exposure was 200 mT for 1 h. Inspired by their impressive results, we decided to expose maize seeds for 1 h and for the same field intensities that they used and, further, to extend the range of the SMF by also testing 300 and 350 mT.

## Materials and methods

Three independent experiments were conducted in total. In each, 8 cardboard packets, each containing 200 maize seeds (obtained from a local provider), were labelled as A, B, C, D, E, F, G, and H. Seven of them were exposed for 1 h to one of the following static MF intensities: 50, 100, 150, 200, 250, 300, or 350 mT applied using an electromagnet GMW 5403 working with a bipolar current source KEPCO BOP20-20ML and a gap of $4.1 \pm 0.1$ cm. The SMF was measured with a TSH481 Hall probe. The eighth packet of seeds was treated identically as the rest, except for the magnet being disconnected; hence, it was "sham exposed" and used as control. The letters A to H were assigned to the MF exposure (or sham exposure) randomly by the operator of the electromagnet (who did not perform the germination assays) and were revealed to the rest of the authors only after the assays were finished and their data statistically analysed; hence, the study was single-blinded. Background SMF and AC magnetic fields inside the stove where the germinations took place were assessed with 3-axis magnetometers HCM5883L (Honeywell, New Jersey, NY) and FW BELL 4190 (OECO, Milwaukie, OR), respectively.

The 200 seeds of each exposed (or sham-exposed) packet were separated in 4 plastic trays (each with 50 seeds), labelled A1, A2, A3, A4, B1, B2, …, G3, G4, H1, H2, H3, H4, making a total of 32 trays. Seeds were laid on a paper towel soaked with distilled water and each tray was put in a polyethylene plastic bag. Then, the 32 trays were placed in a stove at 30 ºC.

The total length (root + shoot) of plantules were measured at the 7th and 10th days of the assay. Three plantule length measurements were made (randomly) per tray; hence, 12 measurements were done (at day 7 and again at day 10) for each magnetic field treatment in each of the 3 independent experiments (n = 3, N = 36, for each treatment). Germinative energy (GE, percentage of viable seeds at day 4 of the assay) and germinative power (GP, percentage of seeds that came to develop into a normal plantule, determined at day 10) were also assessed. GE and GP were determined for each tray for the 3 experiments (n = 3, N = 12, for each treatment).

The statistical analysis was performed with the RStudio software (RStudio Team 2021) and differences were considered statistically significant at a p-value < 0.05. Firstly, combining box plots and Rosner's and Grubbs' tests we identified 5 outliers in the 10th day plantule length measurements and one for both, the germinative power and germinative energy. All these outliers were removed from the datasets before proceeding with the analysis. Secondly, we applied Shapiro's test for normality: for all endpoints, at least one treatment did not have a normal distribution. Therefore, non-parametric tests were performed: the Kruskal-Wallis' test followed by Dunn's post-hoc test for multiple comparisons (with the Bonferroni correction).

## Results

Figure 1a shows the plantule length measured on the 7th and 10th day. It shows that there was a significant increase with respect to control for all SMF intensities except for 350 mT. At the 7th day increases with respect to control ranged between 61.2 % (at 250 mT) and 108.2 % (at 150 mT); all significant p-values were < 0.01. Similar results were seen for the 10th day, increases ranging from 72.3 % and 108.9 %; all significant p-values were < 0.001. There were also significant differences between different SMF intensities, which are presented in detail in Table 1. The treatment that outstood and differentiated more from control and other treatments was the one of 150 mT, while the one that presented the smallest effect was 250 mT. In between, at 200 mT, the biggest dispersion of data was observed, suggesting a sort of threshold around the SMF intensity. Figure 1b displays our observations of the germinative energy (GE) and the germinative power (GP). Their values were, respectively, roughly at 93 % and 95 % for CTRL and for the magnetically treated seeds: no statistically significant difference was introduced by the magnetic treatment on GE nor on GP.

Besides comparing the treated samples against controls, we decided to make an in-depth analysis of the background fields inside the stove the germinations took place. These fields (static, SMF, and variable, AC), could, eventually, play a role as confounder variables so we tested this by comparing the plantule length at the $10^{th}$ day between different zones of the stove. Figures 2a and 2b sketch the four levels of the stove and the detail of how the 32 trays of each experiment (see Materials and Methods) were located on them (the exact same way in all three independent experiment). The SMF was measured at the 4 levels of the stove, on a 10 x 10 cm grid; values were in the following ranges: 18-178 µT, 24-135 µT, 27-219 µT and 22-103 µT for levels L1 (lowest), L2, L3, and L4 (highest), respectively; the highest values being near the front. Torres et al. also measured the SMF inside their stove and reported values between 20 and 140 µT, and of 4.19 mT near a magnet close to the door (Torres et al., 2019). While this values are 3 orders of magnitude below the ones used for the exposures, experiments with SMF in the range 0-200 µT showed changes in gene expression and the quantity of CAB-protein (chlorophyll a,b-binding protein) in Arabidopsis thaliana (Dhiman and Galland, 2018). However, we found no pattern in the background SMF that would resemble our findings in the analysis of our plantule length data, except that the greater mean for the growth at the front (Fig. 2f) might be related to the higher values of the background SMF. It is worth noting that on close examination of the SMF near the metal bars of the shelves, we found values going from ~175 µT down to ~44 µT in just a few millimeters distance. This means that during germination, seeds with the same treatment and germinated side by side in the same tray, might have been exposed to significantly different SMF. This could be a source of variability in the experiments, but hardly of confounding, since trays of all different treatments and levels were expose to these field inhomogeneities. In the future, it would be desirable to use non-metallic shelves.

In addition to SMF, the AC background field was measured at 9 points (3 x 3 grid) at each level and in two situations: with the heater of the stove on (Fig. 2c), and off (Fig. 2d). In these figures, the numbers represent AC magnetic field in $µT_{rms}$; it is clear that with the heater off the background AC field is rather homogeneous and in the order of tens of $nT_{rms}$ (close to the limit of detection of our magnetometer). In contrast, with the heater on the AC fields were greater by a factor of up to ~300 at level 1, passing from tens of $nT_{rms}$ to several $\mu T_{rms}$, and went weaker as the levels go higher (away from the heater resistance of the stove located below the its floor). These fields are intermittent, since the stove resistance goes on and off in order to maintain the temperature. Could the stronger AC background fields at L1 introduce a difference in germination? In the Figure 2e, the plantule length at the $10^{th}$ day can be compared between the different levels of the stove. Indeed, growth at L1 was 56.0 % (p < 0.0001), 45.8 % (p <0.0001) and 36.9 % (p = 0.0012) lower than at L2, L3, and L4, respectively. The smaller lengths at L1 could be due to the AC fields, but on revealing our blinded labels of the trays, we found that, by chance, in two out of the three experiments the control (untreated) seeds grew at L1, so what Figure 2e shows could be not to the AC background fields, but to the fact controls made the mean length go down. Importantly, the AC background alone cannot account for the 108.9 % maximum difference we found (150 mT vs. control), since L1 vs. control was only 56 % different. In Fig. 2f, the analysis for the different zones of the stove shows that seeds at the center grew differently from the back (18.9 %, p = 0.008), the front (16.4 %, p = 0.035) and the left (20.5 %, p = 0.0036), but not from the right; with the greater lengths occurring at the front (trays closer to the stove door). Again, these differences are smaller than the ones we found between treatments vs. control. Nevertheless, there might have been some contribution from subtle differences in the environment inside the stove. All in all, our analysis of the static and AC background fields suggest that, even if they were of any relevance to the germination process, they do not explain the size of the effects we observed, reassuring that the main (if not only) determining factor in the experiments were the 1 h treatments with SMF between 50 and 350 mT.

## Discussion

The capacity of SMF to affect living organisms has been abundantly demonstrated in single cells and animals (Funk et al. 2009; Wiltschko and Wiltschko 2019; Formicki et al. 2019; Burda et al. 2020) -including humans (Chae et al. 2019; Wang et al. 2019)-, and also in plants (Galland and Pazur 2005).

Many studies have been aimed to revealing the fundamental mechanisms of interaction but, to date, they have not been able to provide a whole picture of how this physical stimulus acts upon the physiology of living organisms. One of the better stablished mechanisms (both theoretically and experimentally), is the so called, Radical Pair Mechanism, which states that a static magnetic field can affect the yield of a chemical reaction involving radical pairs intermediates depending on whether they are on a singlet or a triplet configuration (Hore and Mouritsen 2016). However, the application of this explanation to the effects of SMF on plants has been challenged by the experiments in which a SMF reversal induced gene expression changes in Arabidopsis thaliana (Dhiman and Galland 2018), while the RPM predicts that such reversal should not have any effect. An alternative model that does predict sensitivity to a field reversal has been proposed by Binhi and Prato (Binhi 2016; Binhi and Prato 2017), based on the alteration of the precession of magnetic moments due to the externally applied SMF, but this theory has not received as much attention as the RPM. A third, amply discussed mechanism, is the one based on direct action of SMF on biogenic magnetic nanoparticles (BMNP), i.e., particles with a net magnetic moment synthesized by living cells (as opposed to the ones developed and administered artificially). Winklhofer and Kirschvink provided a rigorous mathematical treatment of it (Winklhofer and Kirschvink 2010). BMNP have been reported in all kinds of cells, from bacteria to humans (Gorobets et al. 2017), including the human brain (Kirschvink et al. 1992), and plants (Theil 1987; Gorobets et al. 2019; Priya et al. 2021), and could be the target of SMF. Among them, ferritins, a ubiquitous family of proteins that sequester and hold iron atoms in cavities of 5 to 8 nm in diameter (Theil 1987; Pekarsky and Spadiut 2020), are currently under intensive study (Liu et al. 2016), although not without controversy regarding the plausibility of the proposed physical mechanisms of interaction (Meister 2016; Barbic 2019). A further mechanism proposed is the realignment of liquid crystals by exposure to a SMF. This idea, put forward more than 50 years ago as a possible mechanism of magnetoreception (Labes 1966), remains a current line of investigation (Guillamat et al. 2016). Moreover, in the context of the present work, it is worth noting that liquid crystal-type assemblies have been obtained from plant cell wall extracts and that the hypothesis that a liquid crystal transient state is formed during cell wall construction in plants has been proposed (Reis et al. 1991). Moreover, it has recently been proposed that during mitosis of both animal and plant cells the mitotic spindle is essentially composed of material in a liquid crystalline state (Lydon 2019). It is suggestive that mT-range SMF have been proven to act on liquid crystals, although of inorganic origin (Wang et al. 2014). The bio-physics of liquid crystals promise to be an active and far reaching area of research in the near future (Hirst and Charras 2017).

Regardless of the primary physico-chemical mechanism that gives origin to the interaction (yet to be fully revealed), there are, as a consequence, physiological consequences triggered. For instance, Vashisth and Nagarajan reported an elevated enzymatic activity, early hydration of membranes, and a greater molecular mobility of bulk and hydration water fractions in maize, which they proposed to be responsible of the enhancement of seeds development (Vashisth and Nagarajan 2010). Their findings were further supported by those of Torres et al., who also worked on maize seeds (with maximum effects at 200 mT for 10 min exposures) and demonstrated a change in the enthalpy of water adsorption, favouring it (Torres et al. 2018). There has also been reported an increased in the uptake of nutrients (Na, K, Mn, Mg, Fe, P, among others) in Snow Peas and Chickpeas seedlings from seeds passed twice through 3.5-136 mT SMF for 5 s (Grewal and Maheshwari 2011). Another key physiological variable that has been reported to be affected by SMF was the production of reactive oxygen species (ROS), e.g., in maize, a 39 % increase of hydroxyl radical production (Shine et al. 2017) and a remarkable 320 % increase of $H_2O_2$ content (Kataria et al. 2017). Kataria et al. also found that SMF (200 mT, 1 h) improved the germination and early growth characteristics of maize (and soybean) under salt stress, reporting an increase in shoot length of up to 44% (Kataria et al. 2017). Aladjadjiyan reported a 25% increase in shoot length (150 mT, 10 min) (Aladjadjiyan 2002). Remarkably, Florez et al. tested several combinations of SMF intensity and duration of the exposure and found a 279.1 % increase in total plat length upon continuous exposure for 10 days (Florez et al. 2007). For the same exposure (200 mT, 1 h) that in our experiments yielded a 108.9 % increase in total plant length, Vashisth and Joshi reported a 66 %, although they found 95.8 % in shoot length (Vashisth and Joshi 2017). Table 2 gathers these aforementioned and several other magnetopriming studies on maize found in the literature.

Overall, a clear majority of the studies report beneficial effects of magnetopriming. SMF seem to accelerate plant metabolism, an idea further reinforced by the opposite observation: fields weaker than the geomagnetic one have been proven to inhibit the growth of primary roots of seedlings, decrease cell proliferation, and slow down the cell cycle (Belyavskaya 2004). Nevertheless, in experiments with field intensities between 50 and 250 mT and exposures of 1, 3, 5, and 7 min, Torres et al. observed both favourable and unfavourable effects depending on the combination of field intensity and duration. The greatest favourable effect, a decrease of the mean germination time (MGT) of 12.4 % with respect to control, was found at 50 mT and 1 min, while a 12.0 % increase of the same endpoint happened at 250 mT and 5 min exposure (Torres et al. 2019). Shine et al. also reported detrimental effects for 200-250 mT and longer periods of time (90-120 min) (Shine et al. 2017). While we did not observe an inhibition of growth, our observation of lower or no-effects for fields greater than 250 mT coincide with the findings by Torres et al. and Shine et al. in that there seems to be an amplitude/duration-window for beneficial effects, which in turn reinforces the need of screening exposure parameters in order to optimize benefits of magnetopriming.

## Conclusion

In this study, we found that a 1 h exposure of maize seeds to static magnetic fields of 50-300 mT remarkably enhanced their growth in a 10-day germination assay by a factor of more than 2 as compared to untreated seeds. Germinative energy and germinative power were not affected by the SMF, presumably because those parameters were already high in the control group (92.8 % and 95.1 %, respectively). We challenged our findings with a thorough analysis of the background fields of the stove where germinations took place, but we concluded that (if any) no determinant confounding occur. Our findings are in line with many other in the literature that support that magnetopriming of maize seeds enhance their germination performance.

**Acknowledgements**

The authors thank Dr. Federico Romá for valuable discussions on the design and execution of this study. Grant sponsors: Universidad Nacional de San Luis, Argentina (PROICO 02-0518 and PROICO 03-2220).

**Authors contributions**

LMF: Conceptualization and germination assays; MID: Conceptualization and exposure of seeds to magnetic fields; MFG: Data curation, statistical analysis and figures; LM: Conceptualization, data curation, statistical analysis and figures, and writing-first draft. All authors reviewed the manuscript and approved it for submission.

## Declarations

The authors declare no competing interests.

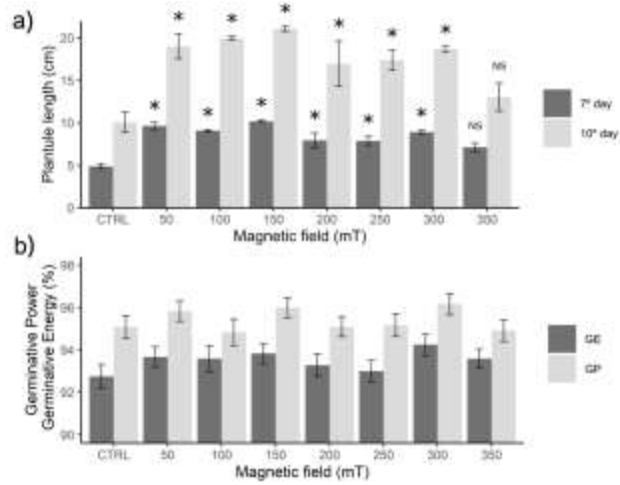

**Fig. 1** a) Plantule length (root + shoot) for the sham-exposed seeds (CTRL) and for several intensities of the magnetic field, from 50 mT to 350 mT at the 7th and 10th day of germination. The star (*) indicates statistically significant difference with respect to CTRL. b) Germinative power (GP) and germinative energy (GE)

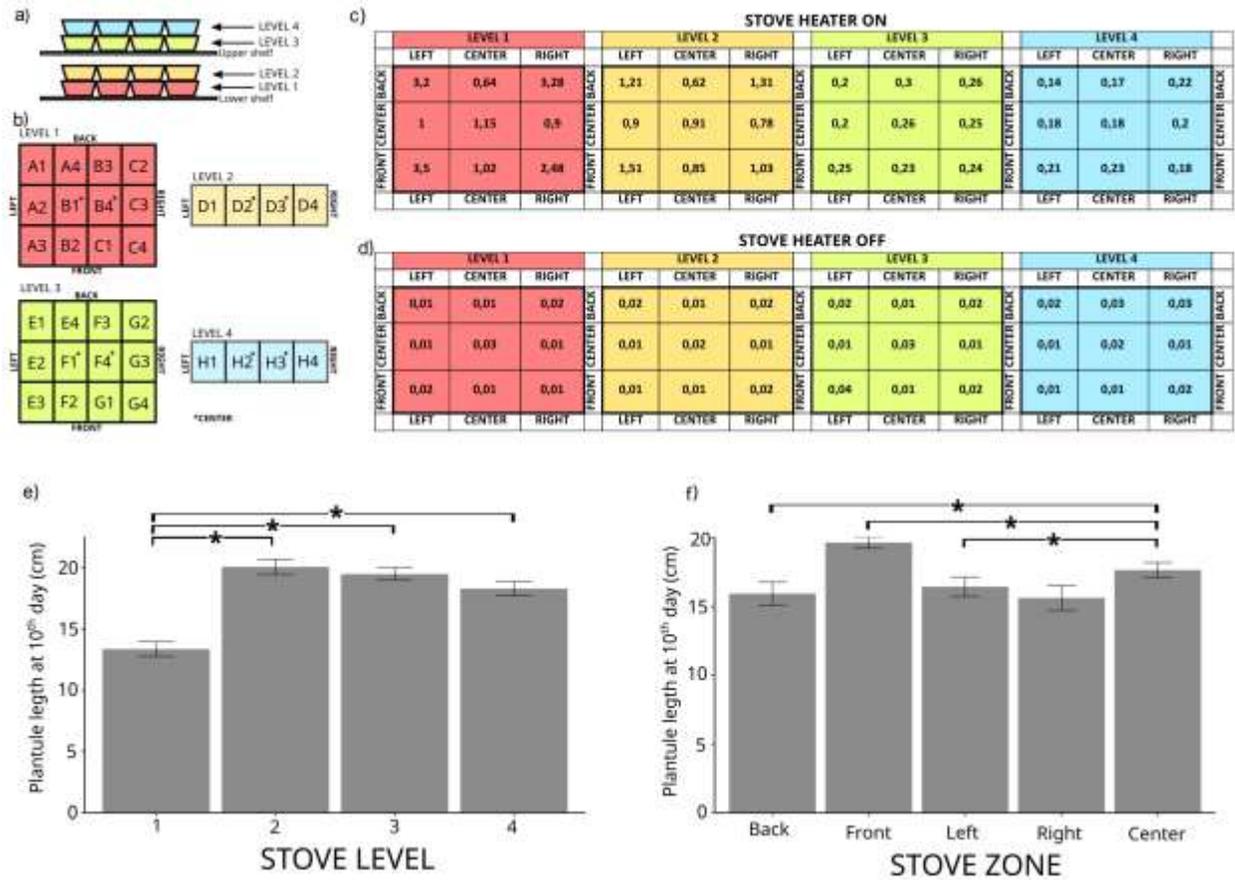

**Fig. 2** a) Each experiment involved 32 trays (4 trays for each of the 8 conditions: 1 control and 7 exposed). 50 seeds were germinated in each of the 32 trays. b) The trays marked with an asterisk (*) were all considered at the center of the stove. c-d) Measurements of AC background fields inside different locations of the stove in which germination assays took place; with its heater on (c) and off (d). All values are in µTRMS. e-f) Plantule length at the 10th day for at the 4 levels (e) and at the different zones (back, front, left, right, and center) of the incubator (f); bars are ±SEM, *: p < 0.05

**Table 1.** Maize seeds response to magnetic field exposure (this study).

| Exposure | Plantule length (cm) 7th day | Plantule length (cm) 10th day | Germinative energy (%) | Germinative power (%) |
|---|---|---|---|---|
| **(a) CTRL** | $4.9 \pm 0.3^{b, c, d, e, f, g}$ | $10.1 \pm 1.2^{b, c, d, e, f, g}$ | $92.8 \pm 0.6$ | $95.1 \pm 0.6$ |
| **(b) 50 mT** | $9.7 \pm 0.4^{a, e, h}$ | $19.0 \pm 1.5^{a, h}$ | $93.7 \pm 0.5$ | $95.8 \pm 0.5$ |
| **(c) 100 mT** | $9.1 \pm 0.1^{a}$ | $20.0 \pm 0.3^{a, h}$ | $93.6 \pm 0.6$ | $94.8 \pm 0.6$ |
| **(d) 150 mT** | $10.2 \pm 0.2^{a, e, f, h}$ | $21.1 \pm 0.3^{a, e, f, h}$ | $93.8 \pm 0.5$ | $96.0 \pm 0.5$ |
| **(e) 200 mT** | $8.0 \pm 0.7^{a, b, d}$ | $17.0 \pm 2.7^{a, d}$ | $93.3 \pm 0.5$ | $95.1 \pm 0.5$ |
| **(f) 250 mT** | $7.9 \pm 0.5^{a, d}$ | $17.4 \pm 1.8^{a, d}$ | $93.0 \pm 0.5$ | $95.2 \pm 0.5$ |
| **(g) 300 mT** | $8.9 \pm 0.2^{a}$ | $18.7 \pm 0.3^{a, h}$ | $94.3 \pm 0.5$ | $96.2 \pm 0.5$ |
| **(h) 350 mT** | $7.1 \pm 0.5^{b, d}$ | $13.1 \pm 1.7^{b, c, d, g}$ | $93.6 \pm 0.5$ | $94.9 \pm 0.5$ |

Data are presented as mean $\pm$ SEM. Differences were considered statistically significant when p-value $< 0.05$ according to Dunn's test (more details in the Statistical analysis section). Superscripts "a", "b", "c",… , and "h" indicate statistical difference with respect to CTRL, 50 mT, 100 mT,… , and 350 mT, respectively (also indicated between parenthesis in the Exposure column). For instance, the plantule length at 10th day exposed to 150 mT ("$21.1 \pm 0.3^{a, e, f, h}$") showed a statistically significant difference with respect to all of the following: (a) CTRL, (e) 200 mT, (f) 250 mT, and (h) 350 mT. For both 7th and 10th day all p-values $< 0.03$.

**Table 2.** Maize seeds response to magnetic field exposure (literature).

| SMF intensity | Exposure duration | Change vs. control (%) | Endpoint* | Reference |
|---|---|---|---|---|
| 150 mT | 10 min | 72 | Dry weight | (Aladjadjiyan 2002) |
| | | 25 | Shoot length | |
| 250 mT | Continuously for 11 days | 24.4 | Fresh weight | (Racuciu et al. 2006) |
| | 1 h | | | |
| 125 mT | 1 h | 12.8 | | |
| 250 mT | Continuously for 10 days | 25.3 | Total plant length | (Florez et al. 2007) |
| 125 mT | | 166.2 | | |
| 250 mT | Continuously for 10 days | 279.1 | | |
| | | 69.2 | Speed of emergence | |
| 560 mT | 30 min | 90.5 | Percentage of establishment | (Domínguez-Pacheco et al. 2010) |
| | | 36.6 | Dry weight | |
| 200 mT | 1 h | 22 | A-Amylase activity | |
| 200 mT | 1 h | 48 | Dehydrogenase activity | (Vashisth and Nagarajan 2010) |
| 100 mT | 2 h | 8 | Protease activity | |
| 100 mT | 2 h | 70.7 | Root-shoot ratio | |
| 200 mT | 1 h | 3.5 | Relative water content | (Anand et al. 2012) |
| 200 mT | 1 h | 20.6 | Leaf water potential | |
| | | 78 | Leaf area | |
| | | 40 | Root length | |
| 200 mT | 1 h | 43 | Superoxide dismutase | (Shine and Guruprasad 2012) |
| | | 23 | Peroxidase | |
| | | 100 | Photosynthetic | |

| | | | performance index | |
|---|---|---|---|---|
| 200 mT | 1 h | 19 44 26 62 320 | Percentage germination Shoot length Root length Water uptake $H_2O_2$ content | (Kataria et al. 2017) (Saline conditions) |
| 125 mT | Continuously for 3.5 days | 16.3 | Mean germination time | (Martinez et al. 2017) |
| 200 mT | 1 h | 20 95.8 49.9 66.0 51.2 97.3 80.6 28.0 | Germination Shoot length Root length Total length Seedling dry weight Seedling vigor I Seedling vigor II Speed of germination | (Vashisth and Joshi 2017) |
| 200 mT | 1 h | 10 58 47 72 22 59 23 39 | Germination Speed of germination Seedling length Fresh weight Dry weight Vigor Index I Vigor Index II Hydroxyl radical production | (Shine et al. 2017) |
| 40 mT | 5 min | 17.9 31.6 | Plant height Leaf Area Index | (Siyami 2018) |
| 50 mT | 1 min | 12.4 | Mean Germination Time | (Torres et al. 2019) |

* Some of the studies reported on many other endpoints, not cited here.